\documentclass[12pt]{iopart}

\usepackage{graphicx}

\begin{document}

\title[Transient state dynamics]
{Neural networks with transient state dynamics}

\author{Claudius Gros}

\address{Institute of Theoretical Physics 
J.W.\ Goethe University Frankfurt, 
60438 Frankfurt, Germany}
\ead{gros07--@--itp.uni-frankfurt.de}
\begin{abstract}
We investigate dynamical systems characterized by a time series
of distinct semi-stable activity patterns, as they are
observed in cortical neural activity patterns. We propose and
discuss a general mechanism allowing for an adiabatic 
continuation between attractor networks and a specific
adjoined transient-state network, which is
strictly dissipative. Dynamical systems with 
transient states retain functionality when their working point
is autoregulated - avoiding prolonged periods of stasis or
drifting into a regime of rapid fluctuations. We show, within
a continuous-time neural network model, that a single local updating rule
for online learning allows simultaneously (a) for information storage
via unsupervised Hebbian-type learning (b) for adaptive regulation of the 
working point and (c) for the suppression of runaway synaptic growth. 
Simulation results are presented, the spontaneous breaking of 
time-reversal symmetry and link symmetry are discussed.
\end{abstract}

\maketitle

\section{Introduction}

Dynamical systems are often classified with respect to
their long-time behaviors, which might be, e.g., chaotic or
regular \cite{katok95}. Of special interest are attractors, 
cycles and limiting cycles, as they determine the fate of all
orbits starting within their respective basins of attraction.

Attractor states play a central role in the theory
of recurrent neural networks, serving the role of memories
with the capability to generalize and to reconstruct
a complete memory from partial initial information \cite{hopfield82}.
Attractor states in recurrent neural networks
face however a fundamental functional
dichotomy, whenever the network is considered as a functional
subunit of an encompassing autonomous information 
processing system, {\it viz} an autonomous cognitive system
\cite{gros07}.
The information processing comes essentially to a standstill 
once the trajectory closes in at one of the attractors. Restarting
the system `by hand' is a viable option for technical
applications of neural networks, but not within the
context of autonomously operating cognitive systems.

\begin{figure}[t]
\centerline{
\includegraphics*[width=0.60\textwidth]{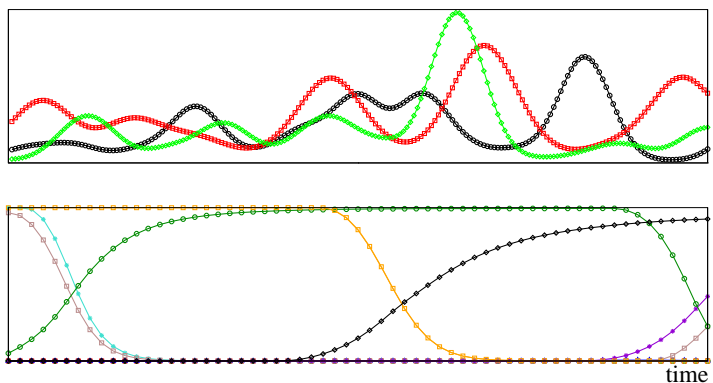}
           }
\caption{Illustration of fluctuating (top) and 
transient state dynamics (bottom).
        }
\label{fig_transStates}
\end{figure}

One obvious way out of this dilemma would be to consider
only dynamical systems without attractor states, i.e.\
with a kind of continously ongoing `fluctuating dynamics', 
as illustrated in Fig.\ \ref{fig_transStates}, which
might possibly be chaotic in the strict sense of
dynamical system theory. The problem is 
then, however, the decision-making process. Without well 
defined states, which last for certain minimal periods, the 
system has no definite information-carrying states onto which 
it could base the generation of its
output signals. It is interesting to note in this context,
that indications for quasi-stationary patterns in cortical 
neural activity have been observed \cite{abeles95,kenet03,ringach03}.
These quasi-stationary states can be
analyzed using multivariate time-series analysis,
indicating self-organized patterns of brain activity
\cite{hutt03}. Interestingly, studies of EEG recordings
have been interpreted in terms of brain states showing
aperiodic evolution states going through sequences of 
attractors that on access support the experience of 
remembering \cite{freeman03}.
These findings suggest that `transient state dynamics', 
as illustrated in Fig.\ \ref{fig_transStates}, might
be of importance for cortical firing patterns.

It is possible, from the viewpoint of dynamical 
system theory, to consider transient states as 
well defined periods when the orbit
approaches an attractor ruin. With a
transient attractor, or attractor ruin,
we denote here a point in phase 
space which could be turned
continously into a stable attractor when 
tuning certain of the parameters entering the
evolution equations of the dynamical system.
The dynamics slows down close to the attractor ruin
and well defined transient states emerge within the ensemble
of dynamical variables. 
The notion of transient state dynamics is related 
conceptually to chaotic itinerancy
\cite{kaneko03}, a term used to characterize
dynamical systems for which
chaotic high-dimensional orbits stay
intermittently close to low-dimensional attractor 
ruins for certain periods. Instability due to 
dynamic interactions or noise is necessary for 
the appearance of chaotic itinerancy.

Having argued that transient-state dynamics
might be of importance for a wide range
of real-world dynamical systems,
the question is then of how to generate
such kind of dynamical behavior in a
controllable fashion and in a manner applicable
to a variety of starting systems. {\it Viz}
we are interested in neural networks which 
generate transient states dynamics in terms 
of a meaningful time series of 
states approaching arbitrarily close
predefined attractor ruins.

The approach we will follow here is to start with an
original attractor neural network and to transform then the 
set of stable attractors into transient attractors
by coupling to auxiliary local variables, which we denote
`reservoirs', governed by long time scales. We note, 
that related issues have been investigated in the context of
discrete-time, phase coupled oscillators \cite{timme02}, 
for networks aimed at language processing in terms of
`latching transitions' \cite{treves05,kropff06}, and
in the context of `winnerless competitions' 
\cite{rabinovich01,seliger03,rabinovich06a}.
Further examples of neural networks capable of generating a 
time-series of subsequent states are
neural networks with time-dependent
asymmetric synaptic strengths \cite{sompolinsky86}
or dynamical thresholds \cite{horn89}. We also
note that the occurrence of spontaneous fluctuating dynamics
has been studied \cite{metzler01}, especially
in relation to the underlying network geometry \cite{paula06}. 

An intrinsic task of neural networks is to learn and to
adapt to incoming stimuli. This implies, for adaptive neural 
networks, a continuous modification of their dynamical 
properties. The learning process could consequently
take the network, if no precautions are taken,
out of its intended working regime, the
regime of transient state dynamics.
Here we will show 
that it is possible to formulate local learning rules
which keep the system in its proper dynamical state
by optimizing continously its own working point. To be
concrete, let us denote with $\bar t$ the average
duration of quasi-stable transient states and with
$\Delta t$ the typical time needed for the transition from 
one quasi-stationary state to the next. The dynamical
working point can then be defined as the ratio $\Delta t/\bar t$. 

These time scales, $\bar t$ and $\Delta t$, result, 
for the network of cortical neurons, from the properties 
of the individual neurons, which are essentially 
time-independent, and from the synaptic strengths, 
which are slow dynamical variables
subject to Hebbian-type learning \cite{arbib02}.
It then follows, that the modifications of the inter-neural
synaptic strengths have a dual functionality: on one
side they are involved in memory storage tasks \cite{arbib02},
and on the other side they need to retain the working point
in the optimal regime. Here we show that this
dual functionality can be achieved within
a generalized neural network model. 
We show that working-point optimization is obtained
when the Hebbian learning-rule is reformulated 
as an optimization procedure, resulting in 
a competition among the set of synapses leading
to an individual neuron. The resulting learning-rule
turns out to be closely related to rules found to
optimize the memory-storage capacity \cite{chechik01}.

\section{Model}

\subsection{Clique encoding}

\begin{figure}[t]
\centerline{
\includegraphics*[width=0.40\textwidth]{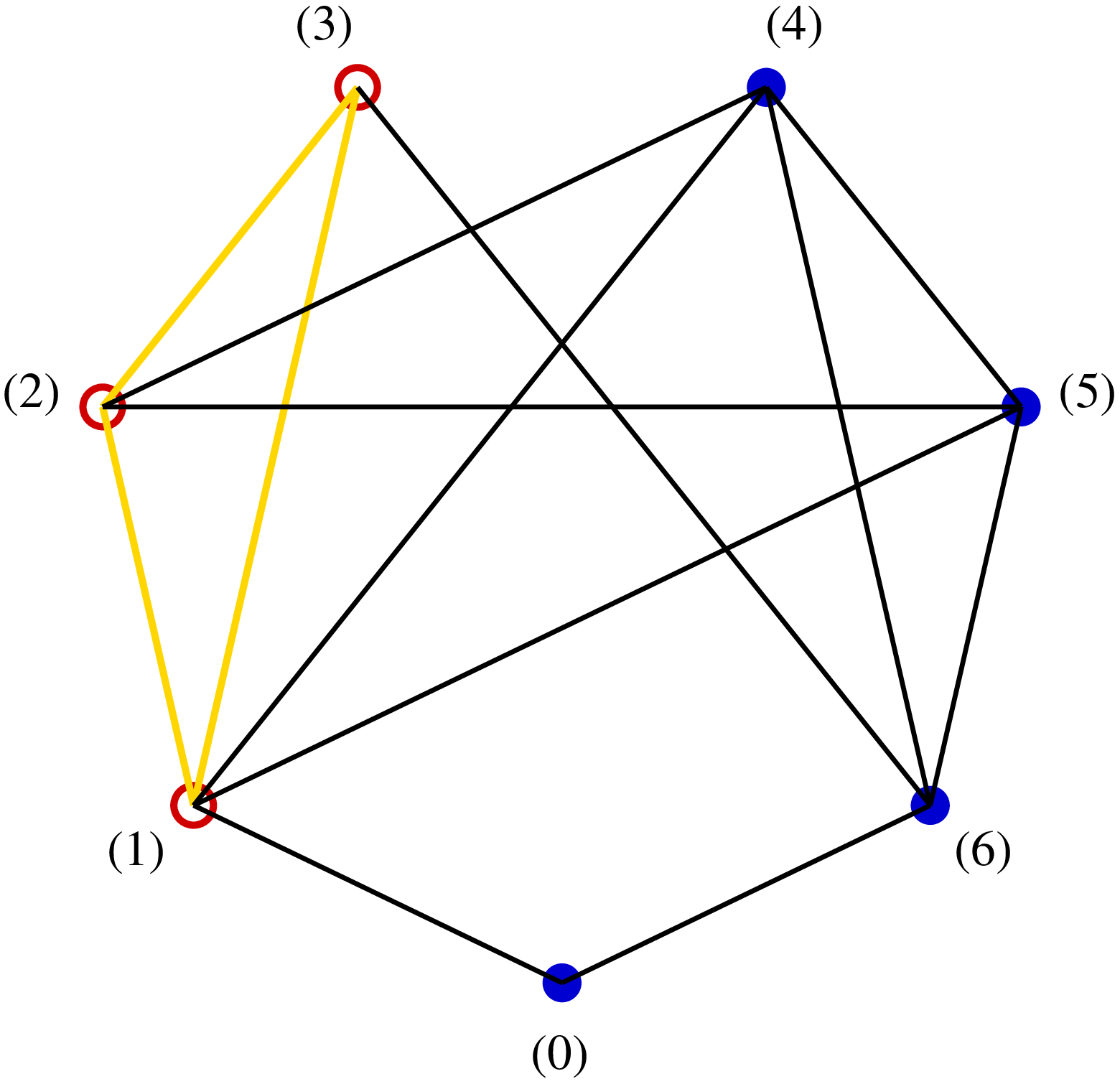}\hspace{1ex}
\includegraphics*[width=0.45\textwidth]{AI7.eps}
           }
\caption{Geometry and simulation results for
a small, 7-site network.\newline
Left: The links with $w_{i,j}>0$, containing six cliques,
(0,1), (0,6), (3,6), (1,2,3) (which is highlighted), 
(4,5,6) and (1,2,4,5). \newline
Right: 
As a function of time,
the activities $x_i(t)$ (solid lines)
and the respective reservoirs $\varphi_i(t)$ (dashed lines)
for the transient state dynamics $(4,5,6)\rightarrow(1,2,3)
\rightarrow (0,6) \rightarrow(1,2,4,5)$. For the 
parameters values see Sect.\ \ref{sec_strict}.
\label{fig_7sites}
        }
\end{figure}

Neural networks with sparse coding, {\it viz} with
low mean firing rates, have very large memory storage
capacities \cite{okada96}. Sparse coding results, in
extremis, in a `one-winner-take-all' configuration, for which
a single unit encodes exactly one memory. In this limit
the storage capacity is, however, reduced again and linearly
proportional to the network size, as in the original
Hopfield model \cite{amit85}.
Here we opt for the intermediate case of `clique encoding'.
A clique is, in terms of graph theory, a fully 
interconnected subgraph, as illustrated in
Fig.\ \ref{fig_7sites} for a 7-site network.
Clique encoding corresponds to a `several-winners-take-all'
setup. All members of the winning clique mutually excite
each other while suppressing the activities of all
out-of-clique neurons to zero.

We note, that the number of cliques can be
very large. For illustration let us consider a random
Erd\"os--R\'enyi graph with $N$ vertices and linking 
probability $p$. The overall number of cliques containing
$Z$ vertices is then statistically given by
\begin{equation}
\left( \begin{array}{c} N \\ Z \end{array} \right)
p^{Z(Z-1)/2}\left(1-p^Z\right)^{N-Z}~,
\label{cogSys_N_z}
\end{equation}
where $p^{Z(Z-1)/2}$ is the probability of having $Z$ 
sites of the graph fully interconnected by $Z(Z-1)/2$ 
edges and where the last term is the probability that 
every single of the $N-Z$ out-of-cliques vertices is 
not simultaneously connected to all $Z$ sites of the clique. 

Networks with clique encoding are especially well suited for 
transient state dynamics, as we will discuss further below,
and are biologically plausible. Extensive sensory preprocessing
is known to occur in the respective cortical areas of the brain
\cite{arbib02}, leading to representations of features and
objects by individual neurons or small cell assemblies. In this 
framework a site, {\it viz} a neural center, 
of the effective neural network considered here
corresponds to such a small cell assembly and a clique to a stable
representation of a memory, by binding together a finite
set of features extracted by the preprocessing
algorithms from the sensory input stream.

\subsection{Continuous time dynamics}

For our study of possible mechanisms of transient state
dynamics in the context of neural networks we consider
$i=1,...,N$ artificial neurons with rate encoding 
$x_i(t)$ and continuous time $t\in [0,\infty]$. Let us
comment shortly on the last point. The majority of research
in the field of artificial neural networks deals with
the case of discrete time $t=0,1,2,...$ \cite{arbib02}.
We are however interested, as discussed in the introduction,
in networks exhibiting autonomously generated dynamical behaviors,
as they typically occur in the context of complete
autonomous cognitive systems. We are therefore interested
in networks having update rules being compatible with
the interaction with other components of a cognitive
system. Discrete time updating is not suitable in
this context, since the resulting dynamical characteristics (i)
depend on the choice of synchronous vs.\ asynchronous
updating and (ii) are strongly influenced when effective
recurrent loops arise due to the coupling to other
components of the autonomous cognitive system. We
therefore consider and study here a model with continuous
time.

\begin{figure}[t]
\centerline{
\includegraphics[height=0.30\textwidth]{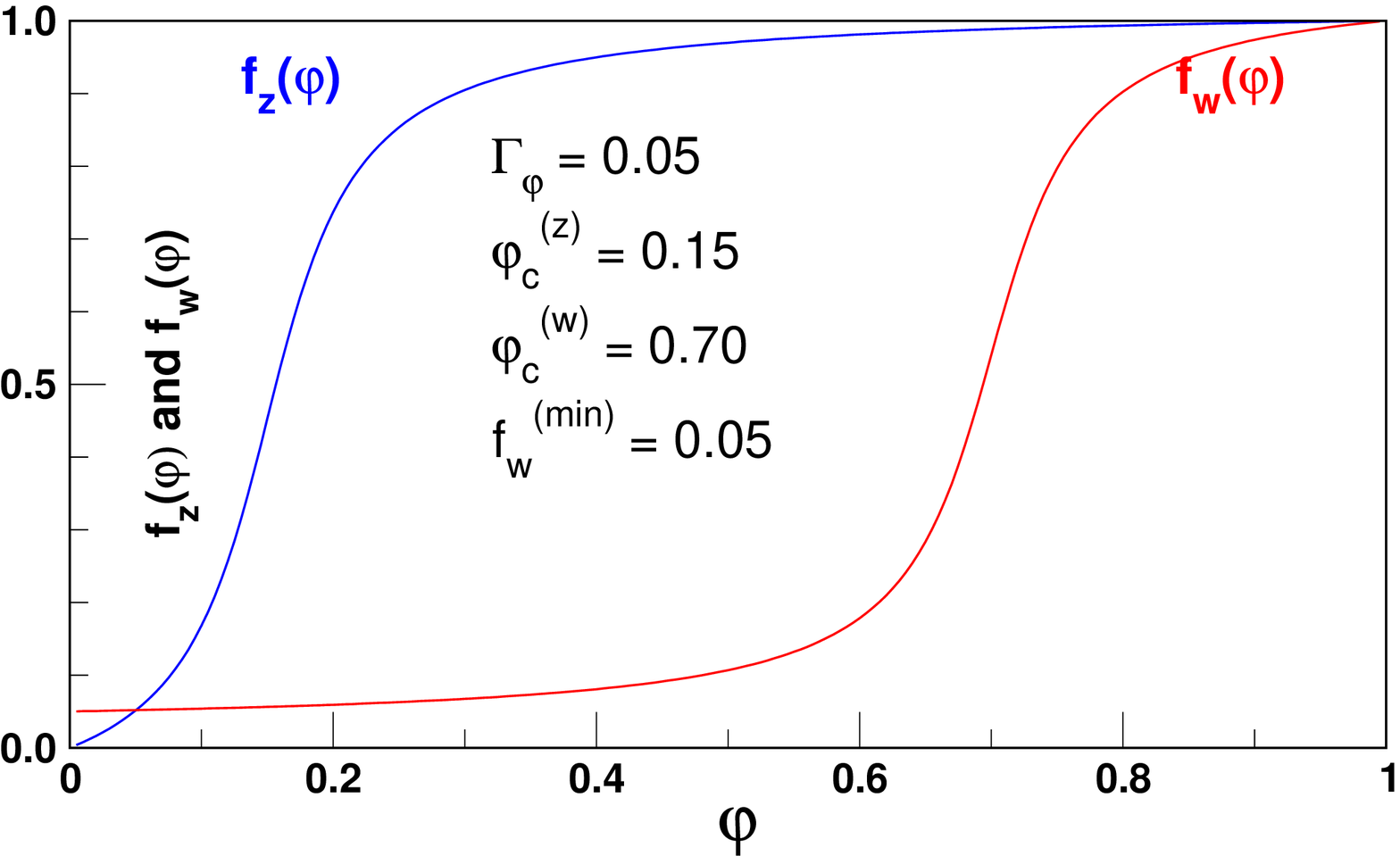}
\hspace{6ex}
\includegraphics[height=0.30\textwidth]{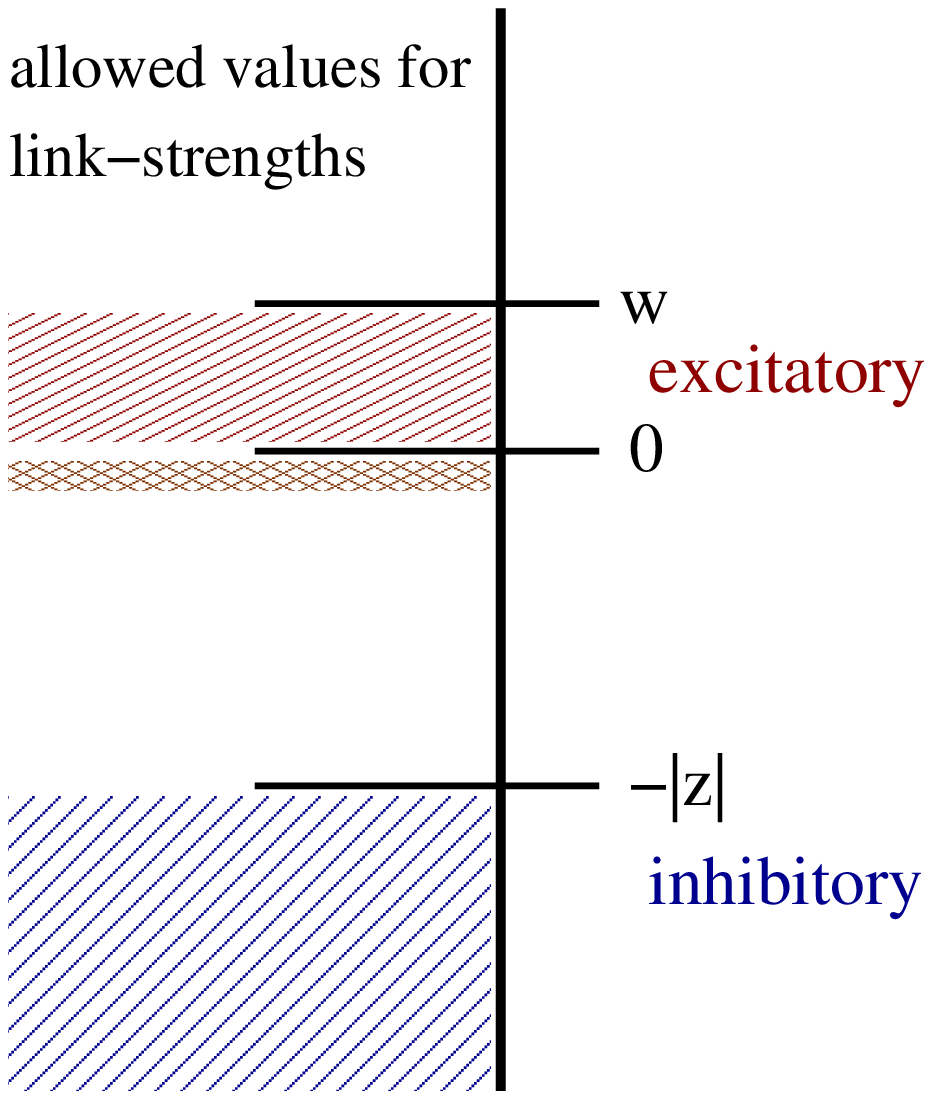}
           }
\caption{Left: Illustration of the reservoir functions
$f_{z/w}(\varphi)$, see Eq.\ \ref{cogSys_ri},
of sigmoidal form \cite{note_functions} 
with respective turning points
$\varphi_c^{(f/z)}$, a width $\Gamma_\varphi$ and a minimal
value $f_z^{(min)}=0$. \newline
Right: Distribution of the synaptic strength for the
inhibitory links $z_{ij}<-|z|$ and the active
excitatory links $0<w_{ij}<w$ leading to
clique encoding. Note, that $w$ is not
a strict upper bound, due to the optimization
procedure (\ref{cogSys_w_L_dot_opt}).
The shaded area just below zero
is related to the inactive $w_{ij}$, see
Eqs.\ (\ref{cogSys_z_t}) and (\ref{cogSys_w_L_dot_opt}).
}
\label{cogSys_fig_gaps}
\end{figure}

\subsection{Neural network model}

We denote the state variables encoding the
activity level by $x_i(t)$ and assume them
to be continuous variables, $x_i\in[0,1]$.
Additionally, we introduce for every site a
variable $\varphi_i(t)\in[0,1]$, termed `reservoir',
which serves as a fatigue memory facilitating 
the self generated time series of transient states.
We consider the following set of differential
equations:
\begin{eqnarray} \label{cogSys_xdot}
\dot x_i &=& (1-x_i)\,\Theta(r_i)\,r_i \,+\, x_i\,\Theta(-r_i)\,r_i 
\\ \label{cogSys_ri} 
    r_i &=& 
    \sum_{j=1}^N \Big[
    f_w(\varphi_i) \Theta(w_{ij}) w_{i,j}
    + z_{i,j}f_z(\varphi_j)
        \Big] x_j\ \ \
\\ \label{cogSys_phidot}
\dot\varphi_i & =&  
\Gamma_\varphi^+\, (1-\,\varphi_i)(1-x_i/x_c)
\Theta(x_c-x_i)
\,-\, \Gamma_\varphi^-\,\varphi_i\,\Theta(x_i-x_c)
\\ \label{cogSys_z_t}
z_{ij} & =& -|z|\,\Theta(-w_{ij})
\end{eqnarray}
We now discuss some properties of 
(\ref{cogSys_xdot}-\ref{cogSys_z_t}), which are suitably
modified Lotka-Volterra equations:
\begin{itemize}

\item \underline{Normalization}\\
Eqs.\ (\ref{cogSys_xdot}-\ref{cogSys_phidot}) respect the 
normalization $x_i,\varphi_i\in[0,1]$, due to
the prefactors $x_i$,$(1-x_i)$, $\varphi_i$
and $(1-\varphi_i)$
in  Eqs.~(\ref{cogSys_xdot}) and (\ref{cogSys_phidot}),
for the respective growth and depletion processes.
$\Theta(r)$ is the Heaviside-step function: 
$\Theta(r<0)=0$  and $\Theta(r>0)=1$.

\item\underline{Synaptic strength}\\
The synaptic strength is split into an
excitatory contribution $\propto w_{i,j}$ 
and an inhibitory contribution $\propto z_{i,j}$,
with $w_{i,j}$ being the primary variable:
The inhibition $z_{i,j}$ is present only
when the link is not excitatory (\ref{cogSys_z_t}).
We have used $z\equiv-1$, {\it viz} $|z|=1$
throughout the manuscript, which then defines the
inverse reference unit for the time development.

\item\underline{Winners-take-all network}\\
Eqs.~(\ref{cogSys_xdot}) and (\ref{cogSys_ri}) describe,
in the absence of a coupling to the reservoir via 
$f_{z/w}(\varphi)$, a competitive winners-take-all 
neural network with clique encoding.
The system relaxes towards
the next attractor made up of a clique 
of $Z$ sites $(p_1,\dots,p_Z)$ connected via 
excitatory links $w_{p_i,p_j}>0$ ($i,j=1,..,Z$). 

\item\underline{Reservoir functions}\\
The reservoir functions $f_{z/w}(\varphi)\in[0,1]$ 
govern the interaction in between the activity levels $x_i$ 
and the reservoir levels $\varphi_i$. 
They may be chosen as washed out step functions
of sigmoidal form \cite{note_functions},
with a suitable width $\Gamma_{\varphi}$ and
inflection points $\varphi_c^{(w/z)}$, see
Fig.\ \ref{cogSys_fig_gaps}.

\item\underline{Reservoir dynamics}\\
The reservoir levels of the winning clique 
depletes slowly, see Eq.~(\ref{cogSys_phidot})
and Fig.\ \ref{fig_7sites}, and recovers only
once the activity level $x_i$ of a given site has
dropped below $x_c$, which defines a site to be
active when $x_i>x_c$. The factor $(1-x_i/x_c)$ occurring in
the reservoir growth process, see the r.h.s.\ of
(\ref{cogSys_phidot}), serves for a stabilization of the
transition between two subsequent memory 
states. When the activity level $x_i$ of a given center
$i$ drops below $x_c$, it cannot be reactivated immediately;
the reservoir cannot fill up again for $x_i\simeq x_c$,
due to the $(1-x_i/x_c)$ in (\ref{cogSys_phidot}).

\item\underline{Separation of time scales}\\
A separation of time scales is obtained when the
$\Gamma_\varphi^\pm$ are much smaller than the
typical strength of an active excitatory link, 
i.e.\ of a typical $w_{ij}>0$, 
leading to transient state dynamics. 
Once the reservoir of a winning clique
is depleted, it looses, via $f_z(\varphi)$, its ability to 
suppress other sites and the mutual intra-clique
excitation is suppressed via $f_w(\varphi)$. 

\item\underline{Absence of stationary solutions}\\
There are no stationary solutions with
$\dot x_i=0=\dot\varphi$ ($i=1,\dots,N$)
for Eqs.\ (\ref{cogSys_xdot}) and (\ref{cogSys_phidot}),
whenever $\Gamma_\varphi^\pm>0$ do not vanish
and for any non-trivial coupling functions 
$f_{w/z}(\varphi)\in[0,1]$.

When decoupling the activities and the reservoir
by setting $f_{w/z}(\varphi)\equiv 1$ one
obtains stable attractors with
$x_i=1/0$ and $\varphi_i=0/1$ for sites
belonging/not-belonging to the winning clique,
compare Fig.\ \ref{fig_escapeManifold}.
\end{itemize}
In Fig.\ \ref{fig_7sites} the
transient state dynamics resulting from
Eqs.\ (\ref{cogSys_xdot}-\ref{cogSys_z_t}) is illustrated.
Presented in Fig.\ \ref{fig_7sites} are data for the autonomous
dynamics in the absence of external sensory signals, we will
discuss the effect of external stimuli further below.
We present in Fig.\ \ref{fig_7sites} only data for a
very small network, containing seven sites, which can 
be easily represented graphically. We have also performed 
extensive simulations for very large
networks, containing several thousands of sites,
and found stable transient state dynamics. 

\subsection{Role of the reservoir}

The dynamical system discussed here represents in first place
a top-down approach to cognitive systems and a
one-to-one correspondence with cortical structures
is not intended. The setput is however inspired
by biological analogies and we may identify
the sites $i$ of the artificial neural network
described by Eq.\ (\ref{cogSys_xdot})
not with single neurons, but with neural
assemblies or neural centers. The reservoir variables
$\varphi_i(t)$ could therefore be interpreted as
effective fatigue processes taking place in continuously
active neural assemblies, the winning coalisions.

It has been been proposed \cite{malsburg86},
that the neural coding used for the binding
of heterogeneous sensory information in terms
of distinct and recognizable objects might
be temporal in nature. Within this temporal
coding hypothesis, which has been investigated
experimentally \cite{gray89}, neural assemblies
fire in phase, {\it viz} synchronous, when
defining the same object and asynchronous when
encoding different objects. There is a close
relation between objects and memories in
general. An intriguing possibility is therefore
to identify the memories of the transient-state
network investigated in the present approach
with the synchronous firing neurons of the
temporal coding theory. The winning coalition
is characterized by high reservoir levels which
would then correspond to the degree of
synchronization within the temporal encoding
paradigm and the reservoir depletion time
$\sim1/\Gamma_\varphi^-$ 
would correspond to the decoherence 
time of the object binding neurons. 

We note, that this analogy can however
not be carried to far, since synchronization
is at its basis a cooperative effect, the
reservoir levels describing on the other
side single-unit properties. In terms
of a popular physics phrase one might
speak of a `poor man's' approach to
synchronization, via coupling to a fatigue
variable.

\subsection{Dissipative dynamics}

The reason for the observed numerical and dynamical 
robustness can be traced back to its relaxational nature.
For short time scales we can consider the
reservoir variables $\left\{\varphi_i\right\}$ to be approximatively
constant and the system relaxes into the next
clique attractor ruin. Once close to a transient attractor, the
$\left\{x_i\right\}$ are essentially constant, {\it viz} close to
one/zero and the reservoir slowly depletes. The dynamics
is robust against noise, as fluctuations affect only details 
of both relaxational processes, but not their overall behavior.

To be precise we note, that the phase space contracts with respect
to the reservoir variables, namely
$$
\sum_i {\partial \dot\varphi_i\over\partial\varphi_i}
\ =\ -\sum_i\left[\Gamma_\varphi^+(1-x_i/x_c)\Theta(x_c-x_i)
                 +\Gamma_\varphi^-\Theta(x_i-x_c)
            \right] \ \le\ 0,
\quad\qquad \forall x_i\in[0,1]~,
$$
where we have used (\ref{cogSys_phidot}). We note
that the diagonal contributions to the link matrices
vanish, $z_{ii}=0=w_{ii}$, and therefore
$\partial r_i/\partial x_i =0$. The phase space 
contracts consequently also with respect to the
activities,
$$
\sum_i {\partial \dot x_i\over\partial x_i}
\ =\ \sum_i \,\Big[\,\Theta(-r_i) -\Theta(r_i)\, \Big]\, r_i 
 \ \le\ 0~,
$$
where we have used (\ref{cogSys_xdot}). The system is therefore
strictly dissipative, leading to the observed numerically robust
behavior.

\subsection{Strict transient state dynamics}
\label{sec_strict}

The self-generated transient state dynamics shown in
Fig.\ \ref{fig_7sites} exhibits well
characterized plateaus in the $x_i(t)$, since
small values have been used for the
depletion and the growth rate of the reservoir,
$\Gamma_\varphi^-=0.005$ and $\Gamma_\varphi^+=0.015$. 
The simulations
presented in Fig.\ \ref{fig_7sites} were performed 
using $w_{ij}=0.12$ for all non-zero excitatory
interconnections.

We define a dynamical system to have `strict
transient state dynamics' if there exists a set of
control parameters allowing to turn the transient 
states adiabatically into stable attractors.
Eqs.\ (\ref{cogSys_xdot}-\ref{cogSys_z_t})
fulfill this requirements, for 
$\Gamma_\varphi^-\to 0$ the average duration $\bar t$ of the
steady-stated plateaus observed in
Fig.\ \ref{fig_7sites} diverges.

Alternatively, by selecting appropriate values
for $\Gamma_\varphi^-$ and $\Gamma_\varphi^+$, it is possible
to regulate the `speed' of the transient state
dynamics, an important consideration
for applications. For a working cognitive system, such
as the brain, it is enough that the transient states 
are stable just for a certain minimal period needed to
identify the state and to act upon it.
Anything longer would just be a `waste of time'.

\begin{figure}[t]
\centerline{
\includegraphics*[width=0.70\textwidth]{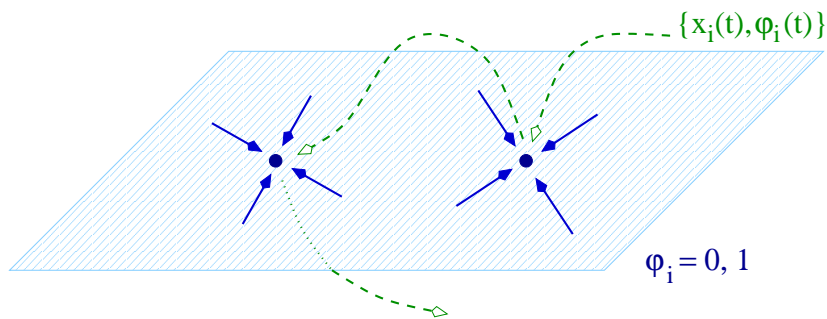}
           }
\caption{The attractors of the original network, 
{\it viz} when the coupling to the reservoir
is turned off by setting $f_{w/z}(\varphi)\equiv 1$,
correspond to
$x_i=1/0$ and $\varphi_i=0/1$ ($i=1,\dots, N$) for
member/non-members of the winning clique. A finite
coupling to the local reservoirs $\varphi_i$
leads to orbit $\left\{x_i(t),\varphi_i(t)\right\}$ 
which are attracted by the attractor ruins for
short time scales and repelled for long time scales.
This is due to a separation of time scales,
as the time evolution of the reservoirs $\varphi_i(t)$
occurs on time scales substantially slower than that of
the primary dynamical variables $x_i(t)$.
        }
\label{fig_escapeManifold}
\end{figure}

\subsection{Universality}

We note that the mechanism for the generation of stable
transient state dynamics proposed here is universal in
the sense that it can be applied to a wide range
of dynamical systems in a frozen state, i.e.\ which are
determined by attractors and cycles.

Physically, the mechanism we propose here is to embed the
phase space $\left\{ x_i\right\}$ 
of an attractor network into a larger space,
$\left\{ x_i,\ \varphi_j\right\}$, by coupling to
additional local slow variables $\varphi_i$. 
Stable attractors are transformed into attractor
ruins since the new variables allow the system to 
escape the basin of the original attractor
$\left\{ x_i=1/0,\ \varphi_j=0/1\right\}$
(for in-clique/out-of-clique sites)
via local escape processes which deplete the respective
reservoir levels $\varphi_i(t)$. Note, that the
embedding is carried out via the reservoir
functions $f_{z/w}(\varphi)$ in Eq.\ (\ref{cogSys_ri})
and that the reservoir variables keep a
slaved dynamics (\ref{cogSys_phidot}) even
when the coupling is turned off by setting
$f_{z/w}(\varphi)\to 1$ in Eq.\ (\ref{cogSys_ri}).

This mechanism is
illustrated in Fig.\ \ref{fig_escapeManifold}.
Locality is an important ingredient for this 
mechanism to work. The trajectories would otherwise 
not come close to any of the attractor ruins again,
{\it viz} to the original attractors,
being repelled by all of them with similar 
strengths and fluctuating dynamics of the
kind illustrated in Fig.\ \ref{fig_transStates}
would result.

\subsection{Cycles and time reversal symmetry}

\begin{figure}[t]
\centerline{
\includegraphics*[width=0.30\textwidth]{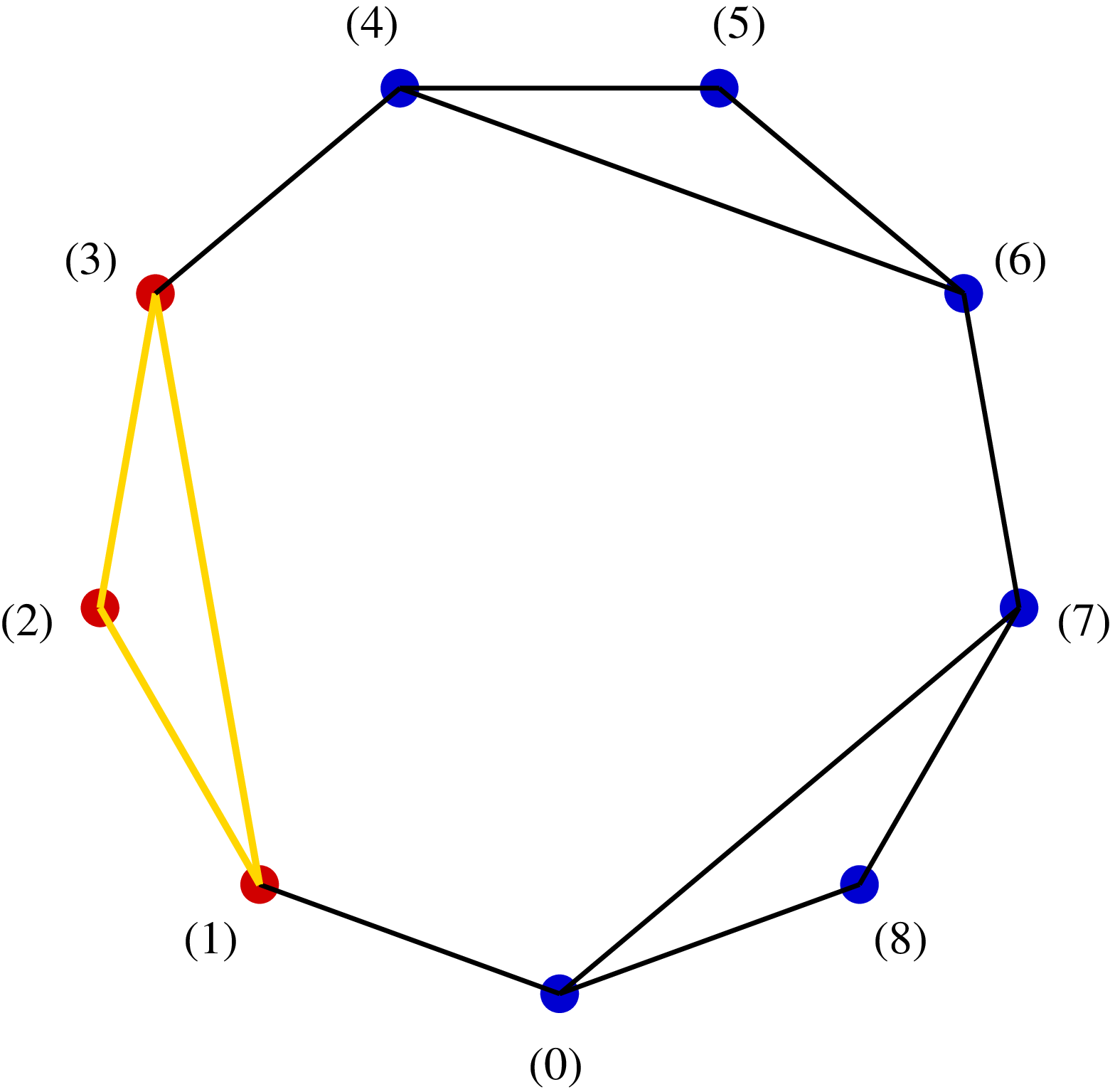}\hspace{1ex}
\lower 3ex \hbox{\includegraphics*[width=0.55\textwidth]{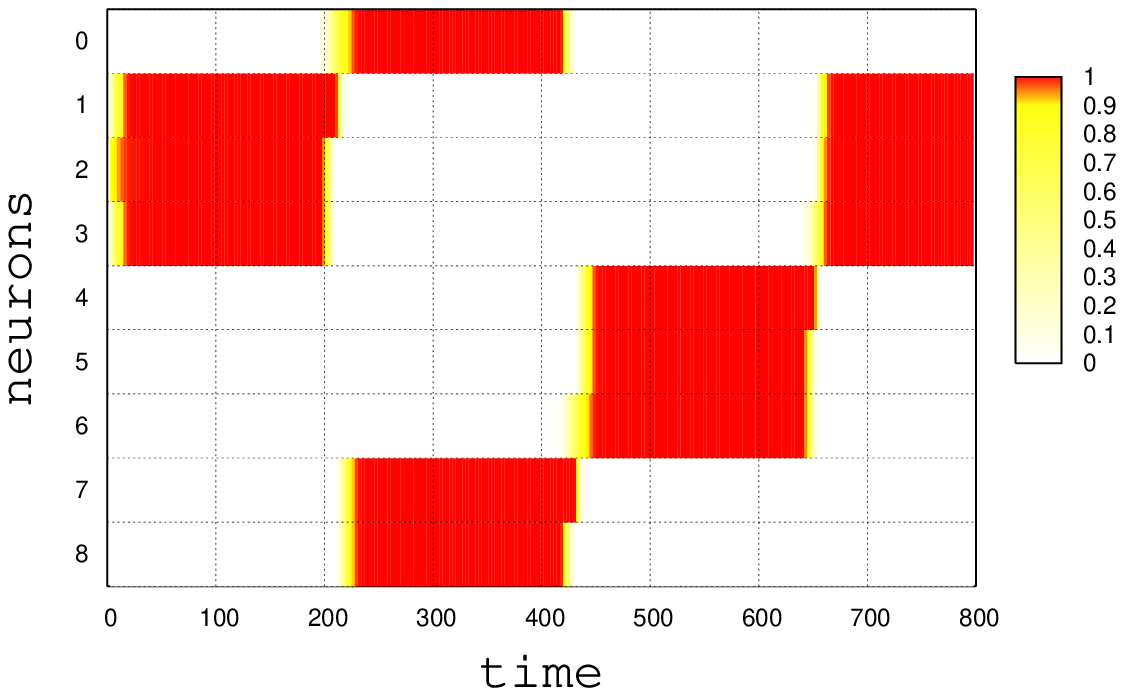}}
           }
\caption{Geometry and simulation results for
a cyclic 9-site network with symmetric excitatory
links $w_{ij}=w_{ji}$.\newline
Left: The links with $w_{i,j}>0$, containing six cliques,
(0,1), (1,2,3) (which is highlighted), 
(3,4), (4,5,6), (6,7) and (7,8,0). \newline
Right: 
As a function of time, the activities $x_i(t)$ 
for the cyclic transient state dynamics $(1,2,3)\rightarrow(7,8,0)
\rightarrow(4,5,6,)\rightarrow\dots$, for the 
parameters values see Sect.\ \ref{sec_strict}. Both
directions (clockwise/anticlockwise) of `rotation' are 
dynamically possible and stable,
the actual direction being determined by the dynamical
initial conditions.
\label{fig_9sites}
        }
\end{figure}

The systems illustrated
in Figs.\ \ref{fig_7sites} and \ref{fig_9sites}
are very small and the transient state dynamics
soon settles into a cycle of attractor ruins, 
since there are no incoming sensory signals 
considered in the respective simulations.
For networks containing a larger number of sites, the
number of attractors can be however very large and such
the resulting cycle length. We performed simulations
for a 100-site network, containing 713 clique-encoded 
memories. We found no cyclic behavior even for sequences
of transient states containing up to 4400 transient states.
We note, that the system does not necessarily retrace its
own trajectory once a given clique is stabilized for a second 
time, an event which needs to occur in any finite system. The
reason being, that the distribution of reservoir levels
is in general different when a given clique is revisited
for a second time.

We note, that time reversal symmetry is `spontaneously'
broken in the sense that repetitive transient state dynamics of
type 
\medskip

\centerline{
(clique A)\ $\to$\ (clique B)\ $\to$\ (clique A)\ $\to$\ (clique B)\ $\to$\ \dots
           }

\medskip\noindent
does generally not arise. The reason is simple. Once the 
first clique is deactivated its respective reservoir 
levels need a certain time to fill up again, 
compare Fig.\ \ref{fig_7sites}. Time reversal symmetry would be
recovered however in the limit 
$\Gamma_\varphi^+ \gg \Gamma_\varphi^-$, 
i.e.\ when the reservoirs would be refilled 
much faster than depleted.

\subsection{Reproducible sequence generation}

Animals need to generate sequences of neural activities
for a wide range of purposes, e.g.\ for movements or for
periodic internal muscle contractions, the heart-beat
being a prime case. These sequences need to
be successions of well defined firing patterns,
usable to control actuators, {\it viz} the muscles.
The question then arrises under which condition a
dynamical system generates reproducible sequences
of well defined activity patterns, i.e.\
controlled time series of transient states \cite{huerta01,rabinovich06b}.

There are two points worth noting in this context.
\begin{description}
\item[i.] The dynamics described by
  Eqs.\ (\ref{cogSys_xdot}-\ref{cogSys_phidot}) 
  works fine for randomly selected link matrices
  $w_{ij}$ which may, or may not  change
  with time passing. In particular one can
  select the cliques specifically in order to induce
  the generation of a specific succession of
  transient states, an example is presented
  in Fig.\ \ref{fig_9sites}. The network is
  capable, as a matter of principle
  to generate robustly large numbers of different
  sequences of transient states. For geometric
  arrangements of the networks sites, and of the
  links $w_{ij}$, one finds waves of transient
  states sweeping through the system.
\item[ii.] In Sect.\ \ref{sec_Autonomous_online_learning}
  we will discuss how appropriate $w_{ij}$ can
  be learned from training patterns presented to
  the network by an external teacher. We will
  concentrate in Sect.\ \ref{sec_Autonomous_online_learning}
  on the training and learning
  of individual memories, {\it viz} of cliques,
  but suitable sequences of training patters could be
  used also for learning temporal sequences of
  memories.
\end{description}

\section{Autonomous online learning}
\label{sec_Autonomous_online_learning}

An external stimulus, $\{b_i^{(ext)}(t)\}$,
influences the activities $x_i(t)$ of the respective neural
centers. This corresponds to a change of the respective
growth rates $r_i$,
\begin{equation}
r_i \ \to \ r_i \,+\,f_z(\varphi_i)\,b_i^{(ext)}(t)~,
\label{cogSys_stim}
\end{equation}
compare Eq.\ (\ref{cogSys_ri}), where $f_z(\varphi_i)$
is an appropriate coupling function, depending on the
local reservoir level $\varphi_i$. When the
effect of the external stimulus is strong, 
namely when $f_z b_i^{(ext)}$ is strong,
it will in general lead to an activation
$x_i\to1$ of the respective neural center $i$.
A continously active stimulus does not convey
new information and should, on the other
hand, lead to habituation, having a reduced 
influence on the system. A strong, continously
present stimulus leads to a prolonged high
activity level $x_i\to1$ of the involved
neural centers, leading via (\ref{cogSys_phidot})
to a depletion of the respective reservoir
levels, on a time scale given by the
inverse reservoir depletion rate,
$1/\Gamma_\varphi^-$.
Habituation is then mediated
by the coupling function $f_z(\varphi_i)$
in (\ref{cogSys_stim}), since
$f_z(\varphi_i)$ becomes very small for
$\varphi_i\to0$, compare
Fig.\ \ref{cogSys_fig_gaps}.
The effect of habituation incorporated in
(\ref{cogSys_stim}) therefore allows the system to 
turn its `attention' to other competing stimuli,
with novel stimuli having a higher chance to affect the
ongoing transient state dynamics.

We now provide a set of learning rules allowing the system
to acquire new patterns on the fly, {\it viz} during
its normal phase of dynamical activity. The alternative,
modeling networks having distinct periods
of learning and of performance, is of widespread
use for technical applications of neural networks, but
is not of interest in our context of continuously active
cognitive systems.

\subsection{Short- and long term synaptic plasticities}

There are two fundamental considerations for the choice
of synaptic plasticities adequate for neural networks with
transient state dynamics.
\begin{itemize}
\item Learning is a very slow process without a short term memory.
      Training patterns need to be presented to the network over and
      over again until substantial synaptic changes are induced \cite{arbib02}.
      A short term memory can speed-up the learning process substantially,
      as it stabilizes external patterns and hence
      gives the system time to consolidate long term
      synaptic plasticity. 

\item Systems using sparse coding are based on a strong
      inhibitory background, the average inhibitory link-strength
      $|z|$ is substantially larger than the average
      excitatory link strength $\bar w$,
$$
  |z|\ \gg\  \bar w~.
$$
      It is then clear that gradual learning is effective only when
      it affects dominantly the excitatory links: Small changes 
      of large parameters do not lead to new transient
      attractors, nor do they influence the cognitive dynamics
      substantially.
\end{itemize}
It then follows, that it is convenient to split the synaptic plasticities
into two parts,
\begin{equation}
w_{ij}\ =\  w_{ij}(t)\ \equiv\  w_{ij}^{S}(t)\,+\, w_{ij}^{L}(t)~,
\label{cogSys_w_S_L}
\end{equation}
where the $w_{ij}^{S/L}$ correspond to the short term and to the
long term synaptic plasticities respectively.

\subsection{Negative baseline}
\label{sec_neg_baseline}

Eq.\ (\ref{cogSys_z_t}), $z_{ij} = -|z|\,\Theta(-w_{ij})$, 
implies that the inhibitory link strength is either
zero or $-|z|$, but is not changed directly during learning,
in accordance to above discussion. 
We may therefore consider two kinds of
`excitatory links strengths':
\begin{itemize}
\item \underline{\sl Active}: An active $w_{i,j}>0$
      is positive and enforces $z_{ij}=0$
      for the same link, via Eq.\ (\ref{cogSys_z_t}).
\item \underline{\sl Inactive}: An inactive $w_{i,j}<0$
      is slightly negative, we use
      $w_{i,j}=W_L^{(min)}<0$ as a default.
      It enforces $z_{ij}=-|z|$
      for the same link, via Eq.\ (\ref{cogSys_z_t})
      and does not contribute to the dynamics,
      since the excitatory links enter as
      $\theta(w_{i,j})$ in (\ref{cogSys_ri}).
\end{itemize}
When $w_{i,j}$ acquires, during learning, a positive value,
the corresponding inhibitory link
$z_{ij}$ is turned off via Eq.\ (\ref{cogSys_z_t})
and the excitatory link $w_{i,j}$ determines
the value of the respective term,
$f_w(\varphi_i) \Theta(w_{ij}) w_{i,j}
    + z_{i,j}f_z(\varphi_j)$,
in Eq.\ (\ref{cogSys_ri}).
We have used a small negative baseline of 
$W_L^{(min)}=-0.01$ throughout the simulations.

\begin{figure}[t]
\centerline{
\includegraphics*[width=0.70\textwidth]{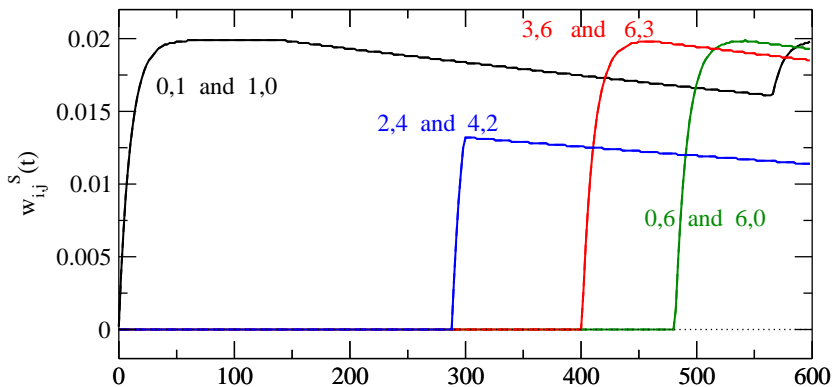}
           }
\caption{The time evolution of the short
term-memory, for some selected
links $w_{i,j}^S$ and the network illustrated
in Fig.\ \ref{fig_7sites}, without the link (3,6).
The transient states are
$(0,1)\rightarrow(4,5,6)
\rightarrow(1,2,3)
\rightarrow(3,6)
\rightarrow(0,6)
\rightarrow(0,1)
$.
An external stimulus at
sites (3) and (6) acts for $t\in[400,410]$ with
strength $b^{(ext)}= 3.6$. 
\label{cogSys_fig_7_AI_learn_S}
        }
\end{figure}

\subsection{Short term memory dynamics}

It is reasonable to have a maximal possible value
$W_S^{(max)}$ for the short term synaptic
plasticities. We consider therefore
the following Hebbian-type learning rule:
\begin{eqnarray}
\label{cogSys_w_S_dot}
\dot w_{ij}^S(t) & =&
\Gamma_{S}^+
\left(W_S^{(max)}-w_{ij}^S\right)
f_z(\varphi_i) f_z(\varphi_j) \,
\Theta(x_i-x_c) \Theta(x_j-x_c) \\
&-& \Gamma_{S}^-\,w_{ij}^S~.
\nonumber
\end{eqnarray}
$w_{ij}^S(t)$ increases rapidly, with rate $\Gamma_S^+$, 
when both the pre- and the post-synaptic neural 
centers are active, {\it viz} when their 
respective activities are above $x_c$.
Otherwise it decays to zero, with a rate $\Gamma_S^-$. 
The coupling functions $f_z(\varphi)$ preempt 
prolonged self-activation of the short term
memory. When the pre- and the post-synaptic centers 
are active long enough to deplete their respective
reservoir levels, the short term memory is shut-off
via the $f_z(\varphi)$. 
We have used $\Gamma_S^+=0.1$, $\Gamma_S^-=0.0005$ and
$W_S^{(max)}=0.02$ and $x_c=0.85$ throughout the simulations.

In Fig.\ \ref{cogSys_fig_7_AI_learn_S} we present the time evolution
of some selected $w_{ij}^S(t)$, for a simulation using the
network illustrated in Fig.\  \ref{fig_7sites}. The short term
memory is activated in three cases: 
\begin{itemize}
\item When an existing clique, {\it viz} a clique encoded in
      the long term memory $w_{ij}^L$, is activated, 
      as it is the case of (0,1) for the data presented in 
      Fig.\ \ref{cogSys_fig_7_AI_learn_S}, the respective
      intra-clique  $w_{ij}^S$ are also activated. This 
      behavior is a side effect since, for the parameter values
      chosen here, the magnitude of the short term link strengths
      is substantially smaller than those of the long term
      link strengths.
\item During the transient state dynamics there is a certain
      overlapping of a currently active clique with the subsequent 
      active clique. For this short time span the short term plasticities
      $w_{ij}^S$ for synapses linking these two cliques get
      activated. An example is the link (2,4) for the simulation
      presented in Fig.\ \ref{cogSys_fig_7_AI_learn_S}.
\item When external stimuli act on two sites not connected by
      an excitatory long term memory link $w_{ij}^L$, the
      short term plasticity $w_{ij}^S$ makes a qualitative difference.
      It transiently stabilizes the corresponding link and the
      respective link becomes a new clique $(i,j)$ either by itself,
      or as part of an enlarged and already existing clique. An example 
      is the link (3,6) for the simulation
      presented in Fig.\ \ref{cogSys_fig_7_AI_learn_S}. Note however that,
      without subsequent transferal into the long term memory, these
      new states would disappear with a rate $\Gamma_S^-$ once the
      causing external stimulus is gone.
\end{itemize}
The last point is the one of central importance, 
as it allows for the temporal stabilization of new 
patterns present in the sensory input stream.

\begin{figure}[t]
\centerline{
\includegraphics*[width=0.70\textwidth]{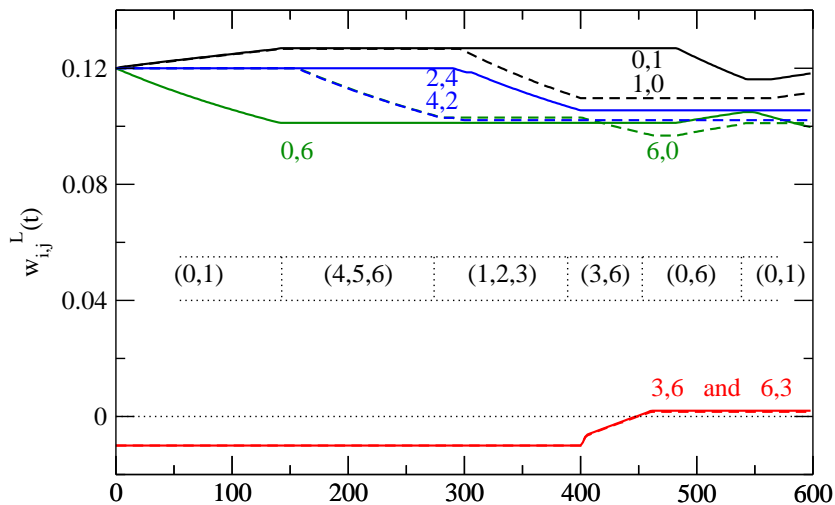}
           }
\caption{The time evolution of the long term memory, 
for some selected links $w_{i,j}^L$ and the network illustrated
in Fig.\ \ref{fig_7sites}, without the link (3,6).
The transient states are
$(0,1)\rightarrow(4,5,6)
\rightarrow(1,2,3)
\rightarrow(3,6)
\rightarrow(0,6)
\rightarrow(0,1)
$.
An external stimulus at
sites (3) and (6) acts for $t\in[400,410]$ with
strength $b^{(ext)}= 3.6$. The stimulus pattern
(3,6) has been learned by the system, as the
$w_{3,6}$ and $w_{6,3}$ turned positive during the
learning-interval $\approx [400,460]$. The learning
interval is substantially longer than the bare
stimulus length due to the activation of the short term
memory. The decay of certain $w_{ij}^L$ in the
absence of an external stimulus is due to forgetting
(\ref{cogSys_w_L_dot_opt}), which should normally
be a very weak effect, but which has been choose here
to be a sizeable $\Gamma_L^-=0.1$,
for illustrational purposes 
 }
\label{cogSys_fig_7_AI_learn_L}
\end{figure}

\subsection{Long term memory dynamics}

Information processing dynamical systems retain 
their functionalities only when they keep their 
dynamical properties within certain regimes, 
they need to regulate their own working point. 
For the type of systems discussed here, exhibiting
transient-state dynamics, the working point
is, as discussed in the introduction,
defined as the time $\Delta t$ the system
needs for a transition from one quasi-stationary
state to the subsequent, relative to the length
$\bar t$ of the individual quasi-stationary states,
which is given by $1/\Gamma_\varphi^-$.

The cognitive information
processing within neural networks occurs on
short to intermediate time scales. For these
processes to work well the mean overall synaptic
plasticities, {\it viz} the average strength of the
long term memory links $w_{ij}^{L}$, needs to be 
regulated homeostatically.
The average magnitude of the 
growth rates $r_i$, see Eq.\ (\ref{cogSys_ri}),
determines the time $\Delta t$ needed to complete a transition
from one winning clique to the next transient state.
It therefore constitutes a central quantity regulating
the working point of the system, since
$\bar t\sim 1/\Gamma_\varphi^-$ is fixed, the
reservoir depletion rate $\Gamma_\varphi^-$
is not affected by learning processes which
affect exclusively the inter-neural synaptic strengths.

The bare growth rates $r_i(t)$ are quite strongly 
time dependent, due to the time-dependence of the postsynaptic 
reservoirs entering the reservoir function
$f_w(\varphi_i)$, see Eq.\ (\ref{cogSys_ri}).
The effective incoming synaptic signal strength
\begin{equation}
\tilde r_i \,=\, \sum_{j}\Big[ w_{i,j}
  \,+\, z_{i,j}f_z(\varphi_j)\Big]x_j~, 
\label{cogSys_tilde_r}
\end{equation}
which is independent of the post-synaptic
reservoir $\varphi_i$, is a more convenient local
control parameter. The working point of the cognitive
system is optimal when the effective
incoming signal is, on the average, of comparable
magnitude $r^{(opt)}$ for all sites,
\begin{equation}
\tilde r_i\ \to\ r^{(opt)}~.
\label{cogSys_r_to_r_opt}
\end{equation}
The long term memory has two tasks: To extract and
encode patterns present in the external stimulus,
Eq.\ (\ref{cogSys_stim}), via unsupervised learning
and to keep the working point of the dynamical system 
in its desired range. Both tasks can be achieved by a 
single local learning rule,
\begin{eqnarray}
\label{cogSys_w_L_dot_opt}
\dot w_{ij}^L(t) & =& 
\Gamma_{L}^{(opt)} 
\Delta \tilde r_i \Big[\,
\left(w_{ij}^L-W_L^{(min)}\right) \Theta(-\Delta \tilde r_i) +
\Theta(\Delta \tilde r_i)
                \,\Big] \\ &&\cdot
\,\Theta(x_i-x_c)\,\Theta(x_j-x_c),
\nonumber \\
& - & \Gamma_L^- \, d(w_{ij}^L)\, \Theta(x_i-x_c)\,\Theta(x_c-x_j)~,
\label{cogSys_w_L_dot_decay}
\end{eqnarray}
where $\Delta \tilde r_i\,=\,r^{(opt)}-\tilde r_i$.
For the numerical simulations we used 
$\Gamma_L^{(opt)}=0.0008$,
$W_L^{(min)}=-0.01$ and $r^{(opt)}=0.2$.
We now comment on some properties of these
evolution equations for $w_{ij}^L(t)$:
\begin{itemize}
\item\underline{Hebbian learning}\newline
      The learning rule (\ref{cogSys_w_L_dot_opt})
      is local and of Hebbian type.
      Learning occurs only when the pre- and the post-synaptic
      neuron are active, {\it viz} when their respective
      activity levels are above the threshold $x_c$.
      Weak forgetting, i.e.\ the decay
      of seldom used links is governed by
      (\ref{cogSys_w_L_dot_decay}). The function
      $d(w_{ij}^L)$ determines the functional dependence
      of forgetting on the actual synaptic strength,
      we have used $d(w_{ij}^L)=\theta(w_{ij}^L)w_{ij}^L$
      for simplicity.
\item\underline{Synaptic competition}\newline
      When the effective incoming signal $\tilde r_i$
      is weak/strong, relative to
      the optimal value $r^{(opt)}$, the active links
      are reinforced/weakened, with $W_L^{(min)}$ being the
      minimal value for the $w_{ij}$. The baseline $W_L^{(min)}$
      is slightly negative, compare
      Figs.\ \ref{cogSys_fig_gaps} and \ref{cogSys_fig_7_AI_learn_L}.
    
      The Hebbian-type learning then takes place in the form of 
      a temporal competition among incoming synapses - frequently 
      active incoming links will gain strength, on the average, 
      on the expense of rarely used links. 
\item\underline{Fast learning of new patterns}\newline
      In Fig.\ \ref{cogSys_fig_7_AI_learn_L} the time evolution
      of some selected $w_{ij}^L$ is presented. 
      A simple input pattern is learned by the
      network. In this simulation the learning parameter
      $\Gamma_{L}^{(opt)}$ has been set to a quite large value
      such that the learning occurs in one step (fast learning).
\item\underline{Suppression of runaway synaptic growth}\newline
      When a neural network is exposed repeatedly to
      the same, or to similar external stimuli, unsupervised
      learning generally then leads to uncontrolled growth
      of the involved synaptic strengths. This phenomena,
      termed `runaway synaptic growth' can also occur
      in networks with continuous self-generated activities,
      when similar activity patterns are auto-generated
      over and over again. Both kinds of synaptic runaway-growth
      is suppressed by the proposed link-dynamics (\ref{cogSys_w_L_dot_opt}).
\item\underline{Negative baseline}\newline
      Note that $w_{ij}=w_{ij}^S+w_{ij}^L$
      enters the evolution equation (\ref{cogSys_ri})
      as $\theta(w_{ij})$. We can therefore distinguish
      between active ($w_{ij}>0$) and inactive
      ($w_{ij}<0$) configuration, compare 
      Fig.\ \ref{cogSys_fig_gaps}. The negative
      baseline $W_L^{(min)}<0$ entering
      (\ref{cogSys_w_L_dot_opt}) then allows 
      for the removal of positive links and
      provides a barrier against small random
      fluctuations, compare Sect.\ \ref{sec_neg_baseline}.
\end{itemize}
During a transient state we have $x_i\to1$ for all vertices
belonging to the winning coalition and $x_j\to0$ for
all out-of-clique sites, leading to
$$
\tilde r_i\ \approx\ \sum_{j}\ w_{i,j}, \qquad\quad j\in\mbox{active\ sites}~,
$$
compare Eq.\ (\ref{cogSys_tilde_r}). The working-point optimization
rule (\ref{cogSys_r_to_r_opt}), 
$ \tilde r_i \to r^{(opt)}$ is therefore equivalent to a local
normalization condition enforcing the sum of active incoming link-strengths
to be constant, i.e.\ site independent. This rule is closely related
to a mechanism of self regulation of the average firing rate of cortical
neurons proposed by Bienenstock, Cooper and Munro \cite{bienenstock82}.

\subsection{Online learning}
\label{subsec_online_learning}

The neural network we consider here is continously active,
independently of whether there is sensory input via
Eq.\ (\ref{cogSys_stim}) or not. The reason being,
that the evolution equations (\ref{cogSys_xdot})
and (\ref{cogSys_phidot}) generate a never ending
time series of transient states. It is a central
assumption of the present study, that continuous and
self-generated neural activity is a condition
sine qua no for modeling overall brain activity
or for developing autonomous cognitive systems
\cite{gros07}.

The evolution equations for the synaptic
plasticities, namely (\ref{cogSys_w_S_dot}) 
for the short-term memory and 
(\ref{cogSys_w_L_dot_opt}) for the
long-term memory are part of the dynamical
system, {\it viz} they determine the time
evolution of $w_{ij}^S(t)$ and of 
$w_{ij}^L(t)$ at all times, irrespectively of
whether external stimuli are presented
to the network via (\ref{cogSys_stim})
or not. The evolution equations for the
synaptic plasticities need therefore
to fulfill, quite in general for a
continously active neural network,
two conditions:
\begin{description}

\item[(a)] Under training conditions, namely when
           input patterns are presented to the
           system via (\ref{cogSys_stim}), the
           system should be able to modify the
           synaptic link strength accordingly,
           such that the training patterns
           are stored in the form of new memories,
           {\it viz} cliques representing 
           attractor ruins and leading to quasistationary
           states.
\item[(b)] In the absence of input the
           ongoing transient-state dynamics
           will lead constantly to synaptic 
           modifications, via  (\ref{cogSys_w_S_dot}) 
           and (\ref{cogSys_w_L_dot_opt}). Theses
           modification may not induce
           qualitative changes, such as the
           the autonomous destruction of existing
           memories or the spontaneous generation
           of spurious new memories. New memories
           should be acquired exclusively via
           training by external stimuli.
\end{description}
In the following section we will present simulations
in order to investigate these points. We find
that the evolution equations formulated in this
study conform with both conditions (a) and (b) above,
due to the optimization principle for the
long-term synaptic plasticities in
Eq.\ (\ref{cogSys_w_L_dot_opt}).

\begin{table}[b]
\caption{Learning results 
for systems with $N$ sites and
$N_{links}$ excitatory links and
$N_2,..,N_6$ cliques containing $2,..,6$ sites.
$N_{tot}$ is the total number memories to
be learned. $N_l$ and $N_{par}$ denote
the number of memories learned completely/partially.
       }
\begin{center}
\begin{tabular}{cc|cccccc|cc}
\hline
$ \ N     \ $ & $ \ N_{links} \  $ & 
$ \ N_2   \ $ & $ \ N_3       \  $ &
$ \ N_4   \ $ & $ \ N_5       \  $ &
$ \ N_6   \ $ & $ \ N_{tot}   \  $ &
$ \ N_{l} \ $ & $ \ N_{par}   \  $ \\
\hline
  20 & 104 &  1 &  10 &  42 & 11 & 1 &  65 &  60 & 3 \\
 100 & 901 & 26 & 563 & 122 &  2 & 0 & 713 & 704 & 7 \\
\hline
\end{tabular}
\end{center}
\label{tab_results_learning}
\end{table}

\section{Simulations}

We have performed extensive simulations
of the dynamics of the network with ongoing learning, for
systems with up to several thousands of sites. We found that
the dynamics remains long-term stable even in the presence 
of continuous online learning governed by Eqs.\ (\ref{cogSys_w_S_dot}) 
and (\ref{cogSys_w_L_dot_opt}), exhibiting semi-regular sequences
of winning coalition, as shown
in Figs.\ \ref{fig_7sites}.
The working point is regulated adaptively and no prolonged
periods of stasis or trapped states were observed in
the simulations, neither did periods of rapid or uncontrolled
oscillations occur.

Any system with a finite number of sites $N$ and a finite
number of cliques settles in the end, in the absence of 
external signals, into a cyclic series of transient states.
Preliminary investigations of systems with $N\approx 20-100$
resulted in cycles spanning on the average a finite fraction of
the set of all cliques encoded by the network. 
This is a notable result, since the overall number 
of cliques stored in the network can easily be
orders of magnitudes larger than the number of sites $N$ itself,
compare Eq.\ (\ref{cogSys_N_z}).
Detailed studies of the cyclic behavior for
autonomous networks will be presented elsewhere.

\subsection{Learning of new memories}

Training patterns $\{p_1,..,p_Z\}$ presented to the system externally via
$r_i \to r_i +f_w(\varphi_i)b_i^{(ext)}(t)$, for
$i\in\{p_1,..,p_Z\}$, are learned by the network
via the activation of the short term memory 
for the corresponding intra-pattern links.
In Figs.\ \ref{cogSys_fig_7_AI_learn_S} and
\ref{cogSys_fig_7_AI_learn_L} we present a case study.
The ongoing internal transient state dynamics is
interrupted at time $t=400$ by an external signal which
activates the short term memory, see Fig.\ \ref{cogSys_fig_7_AI_learn_S}.
Note that the short term memory is activated both
by external stimuli and internally whenever a given
link becomes active, i.e.\ when both pre- and post-synaptic
sites are active coinstantaneous.
The internal activation does however not
lead to the internal generation of spurious memories, since
internally activated links belong anyhow to one or more 
already existing cliques.

In Fig.\ \ref{cogSys_fig_7_AI_learn_L} we present the 
time development of the respective
long term synaptic modifications, $w_{ij}^L(t)$. The
parameters for learning chosen here allow for fast learning,
the pattern corresponding to the external signal, 
retained temporarily in the short term
memory, is memorized in one step, {\it viz} the corresponding
$w_{36}^L(t)$ becomes positive before the transition to the
next clique takes place. For practical applications smaller 
learning rates might be more suitable, as they allow 
to avoid learning of spurious signals generated by
environmental noise.

In Table \ref{tab_results_learning} we present the results
for the learning of two networks with $N=20$ and $N=100$
from scratch. The initial networks contained only two
connected cliques, in order to allow for a non-trivial initial
transient state dynamics, all other links where inhibitory.
Learning by training and storage of 
the externally presented patterns,
using the same parameters as for Figs.\ \ref{cogSys_fig_7_AI_learn_S}
and \ref{cogSys_fig_7_AI_learn_L}, is nearly perfect. The learning
rate can be chosen over a very wide range, as we tested. Here the
training phase was completed, for the 100-site network, 
by $t=5\cdot 10^4$. Coming back to the discussion in
section \ref{subsec_online_learning}, we then conclude that
the network fulfills the there formulated condition
(a), being able to store efficiently training patterns
as attractor ruins in the form of cliques.

\begin{figure}[t]
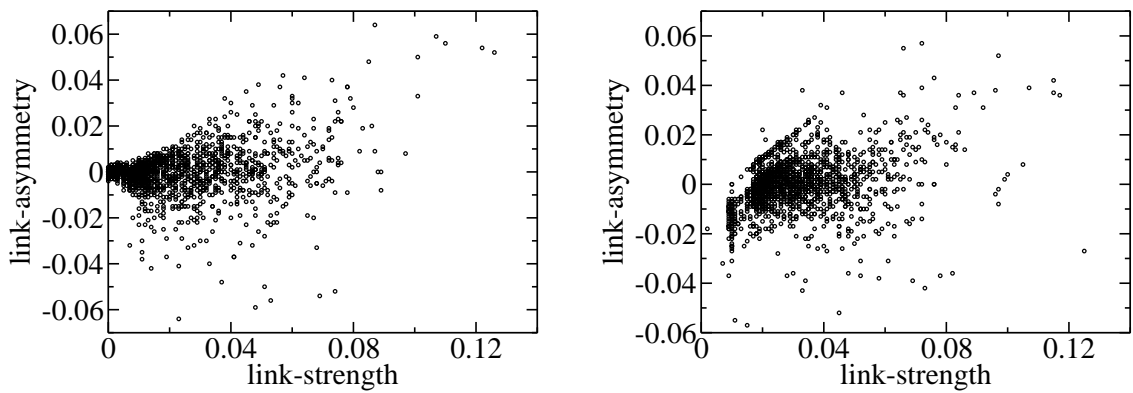

\centerline{
\includegraphics*[width=0.45\textwidth]{100WW_learn.eps}
\hspace{3ex}
\includegraphics*[width=0.45\textwidth]{100WW_start.eps}
           }
\caption{The link-asymmetry $w_{ij}^L-w_{ji}^L$ 
for the positive $w_{ij}^L$ 
for a 100-site network with 713 cliques
at time $t=5\cdot 10^5$, corresponding to circa
4500 transient states.
\newline
Left: After learning from scratch.
Training was finished at $t\approx5\cdot10^4$. \newline
Right: Starting with $w_{i,j}\to0.12$ for
all links belonging to one or more cliques.
\label{cogSys_fig_100_links}
        }
\end{figure}

\subsection{Link asymmetry}

We note that the Hebbian learning via the
working-point optimization, Eq.\ (\ref{cogSys_w_L_dot_opt}),
leads to the spontaneous generation of asymmetries in the
link matrices, {\it viz} to $w_{ij}^L\ne w_{ji}^L$,
since the synaptic plasticity depends on the
postsynaptic growth rates.

In Fig.\ \ref{cogSys_fig_100_links} we present,
for two simulations,  the distribution 
of the link-asymmetry $w_{ij}^L-w_{ji}^L$ 
for all positive $w_{ij}^L$,
for the 100-site network of Table
\ref{tab_results_learning}, at time $t=5\cdot10^5$.
The distributions shown in Fig.\ \ref{cogSys_fig_100_links}
are particular realizations of steady-state distributions,
{\it viz} they did not change appreciably for
wide ranges of total simulation times. 
\begin{description}

\item[(i)] In the first simulation the network had been
learned from scratch. The set of 713 training patterns
were presented to the network for $t\in[0,5\cdot10^4]$.
After that, for $t\in[5\cdot10^4,5\cdot10^5]$
the system evolved freely. A total of 3958 transient 
states had been generated at time $t=5\cdot10^5$,
but the system had nevertheless not yet settled into 
a cycle of transient states, due
to the ongoing synaptic optimization, 
Eqs.\ (\ref{cogSys_w_S_dot}) and (\ref{cogSys_w_L_dot_opt}).
661 cliques remained at $t=5\cdot10^5$, as the link
competition had led to the suppression of some seldom
used links.

\item[(ii)]
In the second simulation, uniform and symmetric
starting excitatory links $w_{ij}^L\to 0.12$ had been set 
by hand at $t=0$, for all intra-clique links. The same $N=100$
network as in (i) was used and the simulation
ran in the absence of external stimuli. All 713 cliques
were still present at $t=5\cdot10^5$, despite the
substantial reorganization of the link-strength
distribution, from the initial uniform to the
stationary distribution shown in Fig.\  \ref{cogSys_fig_100_links}.
A total of 4123 transient states had been generated
in the course of the simulation, without the
system entering into a cycle.

\end{description}
For both simulations all evolution equations,
namely (\ref{cogSys_xdot}) and (\ref{cogSys_phidot})
for the activities and reservoir levels, as well as
(\ref{cogSys_w_S_dot}) for the short-term memory and
(\ref{cogSys_w_L_dot_opt}) and (\ref{cogSys_w_L_dot_decay})
for the long term memory determined the dynamics
for all times $t\in[0,5\cdot 10^5]$. The difference
between (i) and (ii) being the way the memories are
determined, via training by external stimuli,
Eq.\ (\ref{cogSys_stim}), as in (i) 
or by hand as in (ii).

Comparing the two link distributions shown in
Fig.\ \ref{cogSys_fig_100_links}, we note the overall similarity,
a consequence of the continuously acting
working-point optimization. The main differences 
turn-up for small link strengths,
since these two simulations started from opposite
extremes (vanishing/strong initial excitatory links). 
The details of the link distribution shown
in Fig.\ \ref{cogSys_fig_100_links} depend
sensitively on the parameters. For the
results show in Fig.\ \ref{cogSys_fig_100_links}
we used for illustrational purposes
$\Gamma_L^-=0.1$, which is a very big
value for a parameter regulating weak forgetting.
We also performed simulations with $\Gamma_L^-=0$,
the other extreme, and found that the 
link-asymmetry distribution
was somewhat more scattered.

Coming back to the discussion in
section \ref{subsec_online_learning}, we then conclude that
the network fulfills the there formulated condition
(b), since essentially no memories acquired during the
training state were destroyed, or spurious new
memories spontaneously creasted,
during the subsequent free evolution.

\section{Conclusions}

We have investigated a series of issues regarding
neural networks with autonomously generated 
transient state dynamics. We have presented a 
general method allowing to transform an initial
attractor network into a network capable of 
generating an infinite time series of transient
states. The resulting dynamical system has strictly
contracting phase space, with a one-to-one adiabatic 
correspondence between the transient states and 
the attractors of the original network.

We then have discussed the problem of homeostasis,
namely the need for the system to regulate its
own working point adaptively. We formulated a
simple learning rule for unsupervised local Hebbian-type
learning, which solves the homeostasis problem. We note
here, that this rule, Eq.\ (\ref{cogSys_w_L_dot_opt})
is similar to learning rules shown to
optimize the overall storage capacity 
for discrete-time neural networks \cite{chechik01}.

We have studied a continuous time neural network
model using clique encoding and showed that
this model is very suitable for studying
transient state dynamics in conjunction
with ongoing learning-on-the-fly
for a wide range of learning conditions. 
Both fast and slow online learning of new 
memories is compatible with the
transient state dynamics self-generated by
the network.

Finally we turn to the interpretation
of the transient state dynamics.
Examination of a typical time series of 
subsequently activated cliques, as the one
shown in Fig.\ \ref{fig_7sites}, reveals that
the sequence of cliques is not random.
Every single clique is connected to its
predecessor via excitatory links, they are
said to be `associatively' connected \cite{gros05}.
The sequence of subsequently active cliques 
can therefore be viewed, cum grano salis, as
an `associative thought process' \cite{gros05}. The possible
use of such processes for cognitive information
processing needs, however, yet to be investigated.





\begin{thebibliography}{10}
\bibitem{note_functions} The reservoir functions have
the form of generalize Fermi-functions. A possible
mathematical implementation for
$f_\alpha(\varphi)$, with $\alpha=w,z$,
which we used is 
$
f_\alpha(\varphi)\ =\ f_\alpha^{(min)} \,+\, 
\left(1.0-f_\alpha^{(min)}\right)
{
{\rm atan}[(\varphi-\varphi_c^{(\alpha)})/\Gamma_\varphi] -
{\rm atan}[(0-\varphi_c^{(\alpha)})/\Gamma_\varphi]
\over 
{\rm atan}[(1-\varphi_c^{(\alpha)})/\Gamma_\varphi] -
{\rm atan}[(0-\varphi_c^{(\alpha)})/\Gamma_\varphi]
}
$ with $\varphi_c^{(z)}=0.15$, $\varphi_c^{(w)}=0.7$
$\Gamma_\varphi=0.05$, $f_w^{(min)}=0.1$ and $f_z^{(min)}=0$.
\end{thebibliography}
\end{document}